\documentclass{aa}
\bibpunct{(}{)}{;}{a}{}{,} 

\usepackage{amssymb,amsmath,times}
\usepackage{mathtools}
\usepackage{lscape}
\usepackage{epstopdf}
\usepackage{graphicx}
\usepackage{natbib}
\usepackage{booktabs}
\usepackage[normalem]{ulem}

\def\gsim{ \lower .75ex \hbox{$\sim$} \llap{\raise .27ex \hbox{$>$}} }
\def\lsim{ \lower .75ex\hbox{$\sim$} \llap{\raise .27ex \hbox{$<$}} }

\def\beq{\begin{equation}}
\def\eeq{\end{equation}}

\def\sw{{\it Swift}}
\def\fe{{\it Fermi}}
\def\ba{BATSE}

\begin{document}

\title{From the earliest pulses to the latest flares in long GRBs}
\titlerunning{Origin of pulses and flares in GRBs}
\authorrunning{A. Pescalli et al.}
\author{
A. Pescalli\inst{1,3}\thanks{E--mail:alessio.pescalli@brera.inaf.it}, 
M. Ronchi\inst{2},
G. Ghirlanda\inst{3,2},
G. Ghisellini\inst{3,2}
}

\institute{Universit\`a degli Studi dell'Insubria, via Valleggio 11, 22100 Como, Italy.
	\and 
	Dipartimento di Fisica G. Occhialini, Universit\`a di Milano Bicocca, Piazza della Scienza 3, I-20126 Milano, Italy.
	\and
	INAF -- Osservatorio Astronomico di Brera, via E. Bianchi 46, I-23807 Merate, Italy.
    }


\label{firstpage}

\abstract{The prompt emission of Gamma Ray Bursts extends from the early pulses observed in $\gamma$--rays ($>$15 keV) to very late flares of X--ray photons (0.3-10 keV). The duration of prompt $\gamma$--ray pulses is rather constant while the width of X--ray flares correlates with their peak time suggesting a possible different origin. However, pulses and flares have similar spectral properties. 
Considering  internal and external shock scenarios, we derive how the energy and duration of pulses scale with their time of occurrence and we compare with observations. The absence of an observed correlation between prompt emission pulse duration and its time of occurrence favours an ``internal'' origin and confirms the earlier results of Ramirez--Ruiz \& Fenimore. We show that also the energetic and temporal properties of X--ray flares are consistent with being produced by internal shocks between slow fireballs with a small contrast between their bulk Lorentz factors. These results relax the requirement of a long lasting central engine to explain the latest X-ray flares.}


\maketitle



\section{Introduction}
\label{Intro}
Gamma ray bursts (GRBs) are very energetic explosions of $\gamma$ -- rays (prompt emission), detected in the keV--MeV energy range,  characterised by irregular temporal profiles. Lightcurves show variability timescales as short as few milliseconds \citep{Bhat1992Natur.359..217B,Walker2000ApJ...537..264W,MacLachlan2013MNRAS.432..857M},  appearing as a sequence of  ``prompt emission pulses'' (PP herafter) \cite[]{Fishman1993A&AS...97...17F, Bhat2012ApJ...744..141B}. Thanks to the early follow up by the X Ray Telescope (XRT - 0.3--10 keV) onboard Swift \citep[]{Gehrels2004ApJ...611.1005G}, it has been shown that large amplitude ``X--ray flares'' (XRF, hereafter) are often superimposed to the \lq\lq canonical\rq\rq\, afterglow emission \cite[]{Chincarini2007ApJ...671.1903C, Chincarini2010MNRAS.406.2113C, Falcone2007ApJ...671.1921F}. Sometimes, X--ray flares can occur even one day after the $\gamma$--ray trigger \citep{Bernardini2011A&A...526A..27B}.

According to the fireball model, the prompt emission of GRBs is generated by relativistic {\it internal} shocks (IS) produced by shells ejected by the inner engine with random velocities \citep[e.g.][]{Rees1994ApJ...430L..93R}. In these shocks, a fraction of the total kinetic energy of the fireballs is converted into radiation through synchrotron and inverse Compton emission. This scenario can produce the highly variable light curve of the prompt emission \citep{Kobayashi1997ApJ...490...92K}. Shocks produced by the deceleration of the relativistic outflow by the interstellar medium, {\it external} shocks (ES), have been invoked to explain the long lasting, smoothly decaying, broad band (from the optical to the radio) afterglow emission. However, over--densities of the circum burst medium (CBM) could also produce a variable light curve \citep[e.g.][]{Nakar2003ApJ...598..400N}. 

While IS, being produced by shells with slightly different random velocities, are expected to occur at a constant distance from the central engine, in ES the radius where shocks occur increases due to the expansion of the outflow in the CBM. As a consequence, IS should differ from ES in producing pulses whose duration is not correlated with their time of occurrence.
No correlation between the duration and the occurrence time of a pulse of \ba\ GRBs was found \citep{Ramirez2000ApJ...539..712R}. This favoured the IS mechanism. 

\sw\ observed X--ray pulses in the 0.3-10 keV energy range (called flares) which show a duration increasing with time \citep{Chincarini2010MNRAS.406.2113C,Yi2016ApJS..224...20Y,Kocevski2007ApJ...667.1024K}. This property may be consistent with an ES scenario. However, XRFs have spectral properties (e.g. hard to soft evolution and harder spectral shape than the underlying  afterglow component) similar to those of PP \citep{Chincarini2006NCimB.121.1307C,Chincarini2007ApJ...671.1903C,Chincarini2010MNRAS.406.2113C,Falcone2006ApJ...641.1010F,Margutti2010MNRAS.406.2149M} and might be due to IS \citep{Chincarini2007ApJ...671.1903C,Curran2008A&A...487..533C}.  The nature of XRFs is challenging for current models: they might demand a long--lived (hours) central engine \citep[e.g.][]{Yu2015MNRAS.446.3642Y} or they could hint to short lived central engine emitting slower shells which dissipate their energy via IS at later times \citep[e.g.][]{Lazzati2007MNRAS.375L..46L}.  Distinguishing between these two scenarios leads to important implications for the physics of the GRB central engine.

One leading question is whether XRF and prompt emission pulses share the same origin and if they are preferentially produced by IS or ES (the latter due to the interaction with over--densities in the CBM). To answer these questions we derive, under the simplest IS and ES scenarios, the expected relation between the pulse duration and its time of occurrence (\S 2) and compare with observations (\S 3). Discussion and conclusions are presented in \S 4. We assume a standard $\Lambda$CDM flat cosmology with $h=\Omega_{\Lambda}=0.73$. 

\section{Pulse width in internal/external shocks}
\label{PWIE}

IS are thought to be produced by random collisions between shells with different bulk Lorentz factors $\Gamma$. 
Two shells of equal mass $m$ and comparable thickness $\delta R$, moving with Lorentz factors  $\Gamma_{2} > \Gamma_{1} \gg 1$,  collide\footnote{The subscript refer to the spatial ordering of the shell with the faster  ($\Gamma_2$) moving behind the slower one ($\Gamma_1$).} at a radius $R_{\rm c}\simeq \beta_2 c t_{\rm c}$. Here radiation is produced for a timescale comparable to their crossing time. 

Assuming that the two shells are separated by a distance $\Delta R$,
the collision time (as measured by an observer whose viewing angle is $90^\circ$ with respect to the shells' motion direction) is \citep[see][]{Lazzati1999MNRAS.309L..13L}:

\begin{equation} \label{eq:1}
t_{\rm c} = 2 \frac{\Delta R}{c} \frac{\alpha_{\Gamma}^2}{\alpha_{\Gamma}^2-1} \Gamma_1^2
\end{equation}

where $\alpha_{\Gamma}=\Gamma_2/\Gamma_1$. Following \cite{Lazzati1999MNRAS.309L..13L}, we assume that in the collision all the internal (random) energy $\epsilon$ of the merged shell is converted into radiation. This gives an upper limit to the efficiency of conversion of kinetic energy into radiation:
\begin{equation} \label{eq:2}
\eta = 1 - 2\frac{\sqrt{\alpha_{\Gamma}}}{1+\alpha_{\Gamma}} 
\end{equation}
which, under the assumption of shells of equal mass, depends only on the relative ``speed'' of the colliding shells.

After their collision, the merged shells  move with a bulk Lorentz factor $\Gamma_{\rm m} \simeq (\Gamma_1 \Gamma_2)^{1/2}$, at first order approximation. Suppose also that the width of the merged shell does not increase substantially.  An observer located along the direction of the shells' motion will see a pulse whose duration is given by two contributions \citep[e.g.][]{Sari1997ApJ...485..270S,Kocevski2007ApJ...667.1024K}: (i) the ``curvature'' timescale $t_{\rm curv}$ due to the different travel paths of photons emitted simultaneously by the spherical surface\footnote{A spherical shell with an angular scale larger than $\Gamma_{\rm m}^{-1}$ has been assumed.}; (ii) the ``merging'' timescale $t_{\rm merge}$, i.e. the difference of arrival times of photons, emitted along the line of sight, during the time necessary for the two shell to cross one another (also accounting for the relativistic Doppler effect). Therefore, the duration in the rest frame of the source is:

\begin{equation}\label{eq:delta}
\begin{split}
\Delta t^{\rm rest} & = t_{\rm curv}+t_{\rm merge} \simeq \frac{R_{\rm c}}{2c \Gamma_{\rm m}^2} + \frac{\delta R}{c}\frac{\alpha_{\Gamma}}{\alpha_{\Gamma}^{2}-1} \\
& \simeq \frac{\Delta R + \delta R}{c} \frac{\alpha_\Gamma}{\alpha_\Gamma^2-1}
\end{split}
\end{equation}
According to the standard internal shock model, adopted to explain the GRB prompt emission, the relativistic collisions occur between shells intermittently ejected from the inner engine, with $\alpha_{\Gamma} \gtrsim 2$ \citep[e.g.][]{Rees1994ApJ...430L..93R}. Assuming that the initial separation $\Delta R$ is almost the same, we can see from Eq. \ref{eq:1} that the collision time $t_{\rm c}$ and hence the collision radius $R_{\rm c}$ are approximately constant
 (assuming $\alpha_{\Gamma}=2-5$, $\Gamma_1 =50-200$ and $\Delta R = 10^9$ cm, $R_{\rm c}$ is $10^{13-14}$ cm). Therefore, according to Eq. \ref{eq:delta} the duration of pulses produced by IS
 is constant and should not be correlated with the occurrence time of the peaks in the light curve.

The time when pulses occur in a light curve is related to the trigger time of the detector. Therefore we need to evaluate the delay time between the arrival of the trigger photon (the first one produced during the prompt phase) and the photons produced in subsequent relativistic collisions between shells. To better understand this scenario let's assume that the first $\gamma$--ray photon is produced at time $t_{\rm c,p}$ at a radius $R_{\rm c,p} = \beta c t_{\rm c,p}$. Suppose also that the shells producing the following flare/pulse are ejected with a delay time $\Delta T$ with respect to the first shells. Their collision occurs at a radius $R_{\rm c,f}$ and time $t_{\rm c,f}$. The observer will see the pulse after a time  $(R_{\rm c,p} + (\Delta T + t_{\rm c,f} - t_{\rm c,p}  )c - R_{\rm c,f})/c$. The peak time of the pulse $t_{\rm peak}$ is further increased of $\Delta t^{\rm rest}/2$ (assuming triangular shape). The rest frame peak time is:
\begin{equation}\label{eq:t_peak}
t_{\rm peak}^{\rm rest} \simeq  \Delta T + \frac{\Delta R + \delta R}{2c}\frac{\alpha_{\Gamma}}{\alpha^2_{\Gamma}-1} + \frac{\Delta R}{c}\frac{1}{\alpha^2_{\Gamma}-1}
\end{equation}
In Eq.\ref{eq:t_peak} and onwards we omit the term $R_{\rm c,p} - ct_{\rm c,p}$ since its contribution is negligible.
If we allow $\alpha_{\Gamma}$ to assume values lower than 2, $t_{\rm peak}^{\rm rest}$ (Eq. \ref{eq:t_peak}) and $\Delta t^{\rm rest}$ (Eq. \ref{eq:delta}) increases as $\alpha_{\Gamma}$ approaches unity. This means that shells with a small relative speed will collide later, producing a wider pulse and, according to Eq. \ref{eq:2}, dissipating less energy in the relativistic shock \citep[see also][]{Barraud2005A&A...440..809B}.


In the ES scenario pulses can be produced by collisions between a relativistic fireball expanding into 
extended over--density regions (e.g. \citealt{Dermer2000ApJ...534L.155D}, but see \citealt{Nakar2007MNRAS.380.1744N}), at rest with respect to the central engine. We assume that such CBM ``clumps" are distributed at increasing distances from the central engine in shell--like structures. We also require that the thickness $\Delta L$ of these shell--like structures is much smaller than their extension. The interaction of the relativistic shell with such clumps could produce pulses of external origin. The main difference with respect to the IS case is that the shell now collides with targets (the clumps) at rest so that the dissipation is by far more efficient. If the extension of the clumps is always comparable or grater than the jet aperture, the duration of the pulses will be dominated by the ``curvature'' term. Thus, the pulse duration scales with the collision radius which increases linearly with time ($R = 2\Gamma^2 c t$) as the shell expands into the CBM.  

We consider that as the shell expands into the ISM, it collects mass and progressively slows down. Therefore, the dynamic depends on the density profile of the ISM and it changes from one collision to another one. We assume that the dissipation radius (at least of the order of the deceleration radius) is larger than $\Delta L$. The leading term in the pulse duration is due to the curvature effect.

The ISM density profile is described as $n = n_0 r^{-\alpha}$ ($\alpha \geq 0$) where $n_0$ represents the particle density at some characteristic radius $R_0$ close to the GRB progenitor \citep{Chevalier1999ApJ...520L..29C,Panaitescu2000ApJ...543...66P}. The shell bulk Lorentz factor \citep{Blandford1976PhFl...19.1130B,Nava2013MNRAS.433.2107N} for an adiabatic blast--wave is : 
\begin{equation} \label{eq:3}
\Gamma(r)=\left[\frac{ E_{\rm k} (17-4 \alpha)}{16 \pi n_0 m_{\rm p} c^2} \right]^{1/2} r^{-(3-\alpha )/2} 
\end{equation}
 
Computing the radius $R$ of fireball at a time $t$, we get the expression of the pulse duration:
\begin{equation} \label{eq:4}
\Delta t \simeq \frac{R}{2c\Gamma (R)^2 } \simeq (4-\alpha) t
\end{equation} 

Therefore, considering the ES scenario of a shell decelerated by the CBM where pulses are due to over--densities encountered along its path, the pulse width $\Delta t$ should increase linearly with time. 

In summary, for the prompt phase, internal shock and external shock scenarios predict a pulse duration which should be constant or increasing with time, respectively. So far observations seem to favour the internal shock scenario for the prompt phase. The origin of X-Ray Flares is still controversial. However, as seen above, internal shock appears to be a viable mechanism to produce their temporal and energetic features (see \S 3 and \S 4).

\begin{figure*}
\centering
\includegraphics[width= 0.95\textwidth]{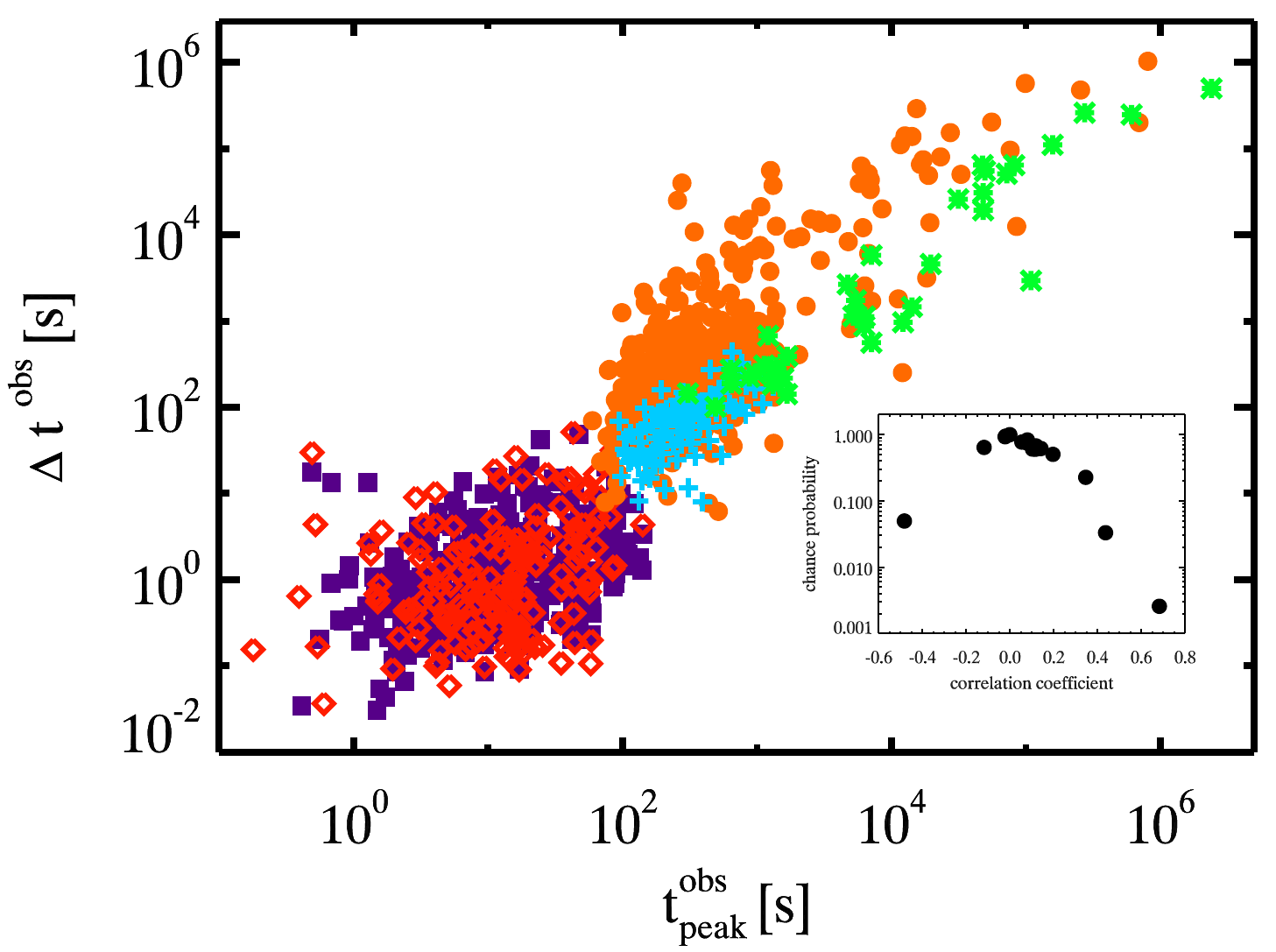}
\caption{\label{observer}{Observer frame pulse duration $\Delta t^{\rm obs}$ versus the pulse occurrence time (peak time - $t_{\rm peak}^{obs}$). Times are refereed to to the trigger time. Prompt emission pulses of \textit{ Fermi}/GBM bursts (from \citeauthor{Bhat2012ApJ...744..141B} \citeyear{Bhat2012ApJ...744..141B}) are shown with different symbols: purple squares show pulses obtained from the analysis of the GBM/NaI [8 keV--1 MeV] lightcurves and red diamonds are pulses from GBM/BGO [200 keV--10 MeV] lightcurves. Orange points, green asterisks and cyan crosses show  X--ray Flares (from \citeauthor{Yi2016ApJS..224...20Y} \citeyear{Yi2016ApJS..224...20Y}, \citeauthor{Bernardini2011A&A...526A..27B} \citeyear{Bernardini2011A&A...526A..27B} and \citeauthor{Chincarini2010MNRAS.406.2113C} \citeyear{Chincarini2010MNRAS.406.2113C}). The bottom--right insert shows the chance probability vs. the correlation coefficient for the 16 GRBs out of 32 in the B12 sample with $\geq 10$ pulses.}}
\end{figure*} 
%

\section{Duration versus peak time}
\label{S}

In this section we compare the simple predictions derived in \S 2 with available observations. 
We consider two samples: the prompt emission pulses (PP) of \fe/GBM bursts recently published by \citet{Bhat2012ApJ...744..141B} and an ensemble of X--Ray Flashes (XRF) detected by \sw/XRT (in the [0.3--10] keV energy range) and published in different papers \citep{Chincarini2010MNRAS.406.2113C,Bernardini2011A&A...526A..27B,Yi2016ApJS..224...20Y}. Below we summarize the main features of these samples. Our aim is to study how the temporal properties (duration and peak time) correlates and how the pulse duration and energetic evolves with the pulse time from the earliest PP to the latest XRF.

For the prompt pulses (PP) we consider the sample of bright GRBs analyzed in \citeauthor{Bhat2012ApJ...744..141B} (\citeyear[B12 hereafter]{Bhat2012ApJ...744..141B}). They selected 32 bright long GRBs (only 7 with measured redshift), detected by \fe\ in its first year of activity, which have the product of their fluence and peak flux larger than $10^{-4}$ erg photons cm$^{-4}$ s$^{-1}$. B12 decompose the observed lightcurves (considering both the data of the NaI [8\,keV--1\,MeV] and BGO [200\,keV--10\,MeV] detectors) as the superposition of log--normal pulses obtaining, for each GRB in their sample, a set of pulse duration, peak time and intensity. 
B12 analyzed light curves with a variable time resolution between 25 and 50 milliseconds in order to maximize the number of fitted pulses. Since we are interested in the pulse duration, we considered only the pulses with duration $>$25 ms, i.e. reasonably larger than the time resolution of the light curves analysed by B12. We extracted from their compilation 374 pulses obtained from the analysis of the light curves of the NaI detectors, and  228 pulses from the data of the BGO detectors.
 
For the XRF there are different samples published in the literature: \citeauthor{Chincarini2010MNRAS.406.2113C} (\citeyear[C10]{Chincarini2010MNRAS.406.2113C}), \citeauthor{Bernardini2011A&A...526A..27B} (\citeyear[B11]{Bernardini2011A&A...526A..27B}), \citeauthor{Yi2016ApJS..224...20Y} (\citeyear[Y16]{Yi2016ApJS..224...20Y}). C10 analyzed 113
early flares ($t_{\rm peak} < 1000$ s) detected by \textit{Swift} between April 2005 and March 2008. B11, extending to flares detected up to December 2009, considered 36 late flares ($t_{\rm peak} > 1000$ s). Joining these two samples, the total number of flares amounts to 149 of which 59 belongs to GRBs with measured redshift. Recently, Y16 enlarged the sample of XRFs considering all GRBs up to March 2015. Their catalog contains 468 bright and significant XRFs of which 200 with redshift. We collected the flares from these three works. The flare durations are estimated differently in these samples: C10 and B11 use the Norris function \citep{Norris2005ApJ...627..324N} to fit flare light--curves and derive the flare width as the difference between the two characteristic e--folding intensity times along the rising/decaying fitted profile.
Therefore, the early and late flares of C10 and B11, sharing the same method, provide a uniform estimate of flare durations. Y16 defines the duration as the difference between the intersection points of the flare fitted profile and the underlying power--law fitted continuum. For the flares in common between C10+B11 and Y16 we verified that the estimated durations are comparable. Moreover, comparing the duration distributions of the sample of C10+B11 and of Y16 we found a KS probability of 0.25 that the duration distributions are drawn from the same parent distribution.

Fig.\ref{observer} shows the (observed) duration $\Delta t^{\rm obs}$ of pulses as a function of the observed time of the pulse peak $t_{\rm peak}^{\rm obs}$. Prompt emission pulses obtained from B12 are shown by the purple squares and red diamonds (corresponding to BGO and NaI data, respectively) and XRFs are shown by the orange points (Y16), green asterisks (B11) and cyan crosses (C10). Peak times are referred to the trigger time of individual GRBs they belong to. 

The distribution of PP and XRF in the plane of Fig.\ref{observer} seems to describe an overall, almost linear, continuum extending from short duration (e.g. 0.01 seconds) early (1 s post trigger) pulses to extremely long and late flares (up to 11 days after the trigger and with comparable duration). We note that a possible selection effect on X--ray flares is due to the time needed for XRT to repoint the GRB. Typically this time is 1 minute which is close to the division between PP and XRF in Fig.\ref{observer}. 

\subsection{Correlation analysis}

We studied the possible correlation between the pulse duration and its peak time in Fig.\ref{observer}. Considering PP and XRF as two distinct populations we find the presence of a correlation (stronger and more significant for flares) between the pulse duration $\Delta t^{\rm obs}$ and the peak time $t_{\rm peak}^{\rm obs}$. The Spearman's correlation coefficient and its chance probability are $r=0.31$ and $P\sim 10^{-15}$, respectively, for prompt pulses (considering NaI and BGO pulses together, i.e. purple squares and red diamonds in Fig.\ref{observer}). For XRF, distributed in a larger region of the $\Delta t^{\rm obs}$--$t_{\rm peak}^{\rm obs}$ plane with respect to PP, we find $r=0.60$ ($P\sim10^{-61}$). We note that the duration and peak time of late--time flares, i.e. with duration $> 10^{4}$ s, might suffer from the difficulty of \textit{Swift}/XRT to continuously follow the flare emission along its orbit. \citet{Bernardini2011A&A...526A..27B} and \citet{Yi2016ApJS..224...20Y} considered only flares with a well covered light--curve (i.e. rise, peak and decay). We verified, however, if the correlation between the flare duration and its peak time holds when conservatively considering only early/intermediate duration flares. We restrict to flares with duration $< 10^{3}$ s and performed the correlation analysis. We find a correlation coefficient of 0.42 and a chance probability of $10^{-23}$.

Since both $\Delta t^{\rm obs}$--$t_{\rm peak}^{\rm obs}$ are computed in the observer frame we verified that the correlations are not induced by the common redshift dependence. 
Since only few
bursts in B12 sample have measured $z$, we performed a Monte Carlo simulation randomly generating redshifts from the GRB formation rate (as reported by \citeauthor{Li2008MNRAS.388.1487L} \citeyear{Li2008MNRAS.388.1487L}; \citeauthor{Pescalli2016A&A...587A..40P} \citeyear{Pescalli2016A&A...587A..40P}). We created $10^4$ random samples for which we computed the correlation coefficient. For XRFs we considered the 259 bursts with measured redshifts. We computed the partial correlation coefficient accounting for the common dependence on $z$ of $\Delta t^{\rm obs}$ and $t_{\rm peak}^{\rm obs}$.  
Also this test shows the correlation found in XRF is solid and not induced by $z$. For PP we again find no significant correlation. 


In order to verify if there is a correlation within individual bursts we can compute the correlation coefficient between the pulse duration and its peak time within single events. However, while for XRF there are only a handful of events with more than 4 flares which prevent the assessment of the significance of the correlation, it is possible to perform such a test with the prompt emission pulses. We considered, within the sample of 32 GRBs of B12, the 16 events with more than 10 pulses. For these we computed individually the correlation coefficient between the duration of the pulses and their time of occurrence. The results are shown in the bottom--right inset of Fig.\ref{observer}. We note that in most cases no significant correlation (i.e. the chance probability is $>$0.1) is present.
These tests suggest that, on average, no correlation is present between the pulse width and its time of appearance in prompt emission pulses (PP) as originally found by \cite{Ramirez2000ApJ...539..712R} while a positive correlation exists in XRF \citep{Chincarini2010MNRAS.406.2113C,Yi2016ApJS..224...20Y,Kocevski2007ApJ...667.1024K}.

\cite{Ramirez2000ApJ...539..712R} considered GRBs detected by \ba, aligned the light curves to the time of the brightest peak during individual bursts and normalized the peak--aligned light curves. However, their analysis missed the possible corrections for the redshift on the energy and time.

In order to verify their results including the corrections for $z$, we select another sample of \fe\ long GRBs ($T_{\rm 90} > 2$ s) with redshift measurement in the \fe\ database (the B12 sample contains only 7/32 GRBs having $z$). We find 100 bursts (from GRB\,080804  to GRB\,160629). We excluded 32 bursts having SNR lower than 3\footnote{SNR was calculated as $\overline{(S-B)}/\bar{B}^{0.5}$, where S is the signal and B the background. The background has been fitted with a polynomial function over two time intervals selected after and before the temporal region containing the event of interest.}, GRB\,130427 which saturated the \fe/GBM detectors and GRB\,120624 because the TTE data (necessary to produce high--resolution light curves) were only partially available due to the instrument slewing~\citep[][GCN \#13377]{Gruber2012GCN.13377....1G}. For the 64 remaining bursts we extracted the light curves in the common $[72,800]$ keV {\it rest frame energy range}\footnote{In this energy range NaI detectors have an almost constant response efficiency.} with 256 ms temporal resolution, from the most illuminated NaI detector. 
The background subtracted light curves are converted to rest frame times and resampled on a common time grid. The resolution of the new time grid has been chosen in order to faithfully reproduce the original lightcurves without losing any information on the time structure and variability.  

\begin{figure}
\centering
\includegraphics[width= 0.45\textwidth]{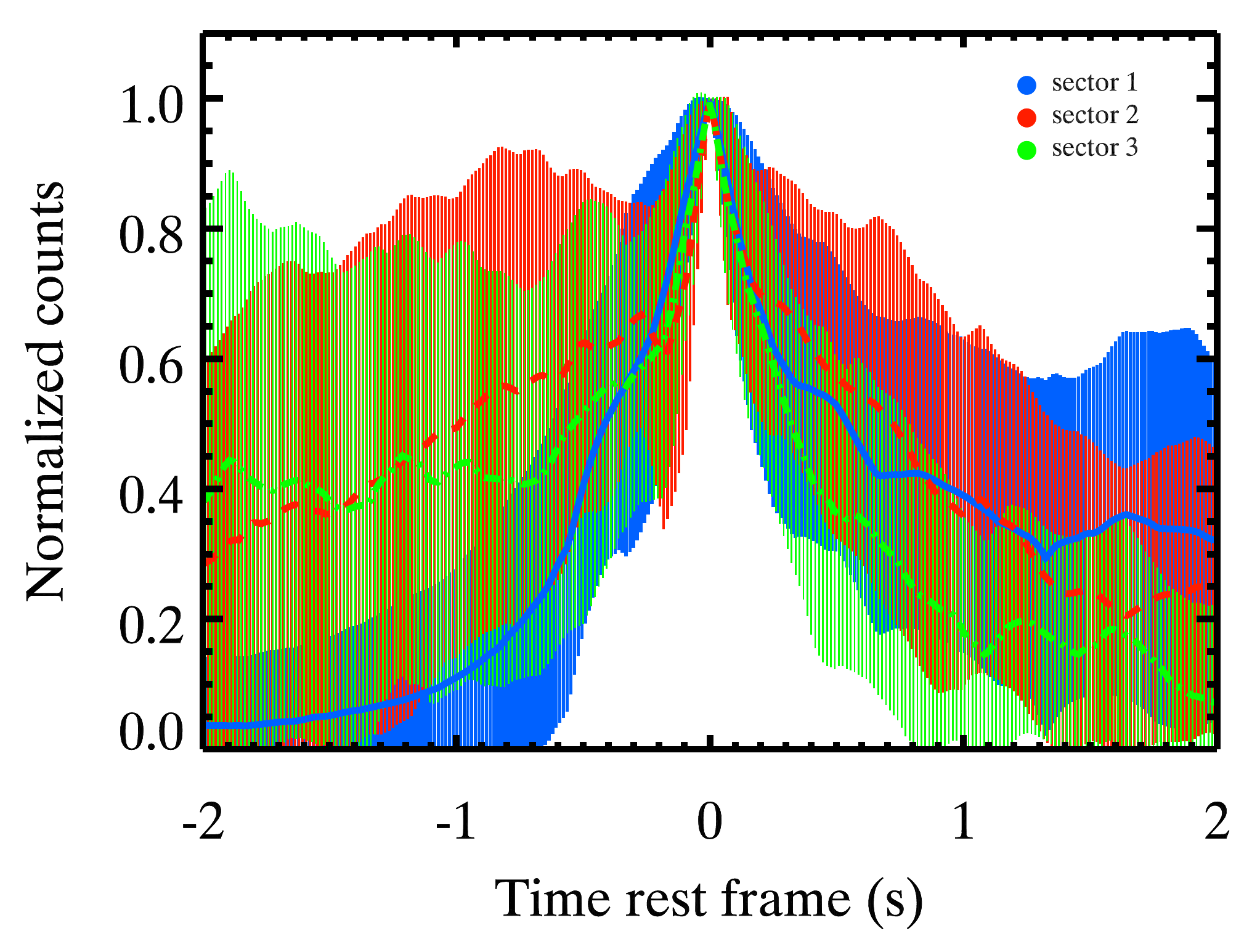}
\caption{\label{profilimedi}{Peak time aligned signal for the three sectors identified within each GRB. Different combinations of colors and line--styles refer to different sectors.}}
\end{figure}

We applied the Average Peak Alignment (APA) method \citep{Mitrofanov1996ApJ...459..570M} dividing light curves in three equal parts (according to the rest frame $T_{90}$ of each bursts). These are the ``sectors'' shown in Fig.\ref{profilimedi} with different colors. The uncertainty on the pulse aligned signal (solid line in Fig.\ref{profilimedi}) is computed as the standard deviation of the signal in each bin. Since secondary peaks are not aligned in time this results in a large dispersion of the curve aside of the main (aligned) peak.
Following \cite{Ramirez2000ApJ...539..712R} we also analysed separately with the APA method  GRBs with $T_{\rm 90} > 20$ s and GRBs with $T_{\rm 90} < 20$ s. Even considering these subsamples, we find similar results to those shown in Fig.\ref{profilimedi}.

By visual inspection of Fig. \ref{profilimedi} we confirm, even accounting for the energy and time redshift corrections, that there is no evolution of the pulse width with time during the prompt emission phase of GRBs. However, the pulse alignment method has some limitations: it allows us to compare preferentially the main emission episodes of different light curves and it is rather less sensitive to the whole emission (and weaker pulses) of the GRB. The wings of the profiles in Fig.\ref{profilimedi} show  the large uncertainties due to the great diversity of temporal profiles.  

\begin{figure}
\centering
\includegraphics[width= 0.45\textwidth]{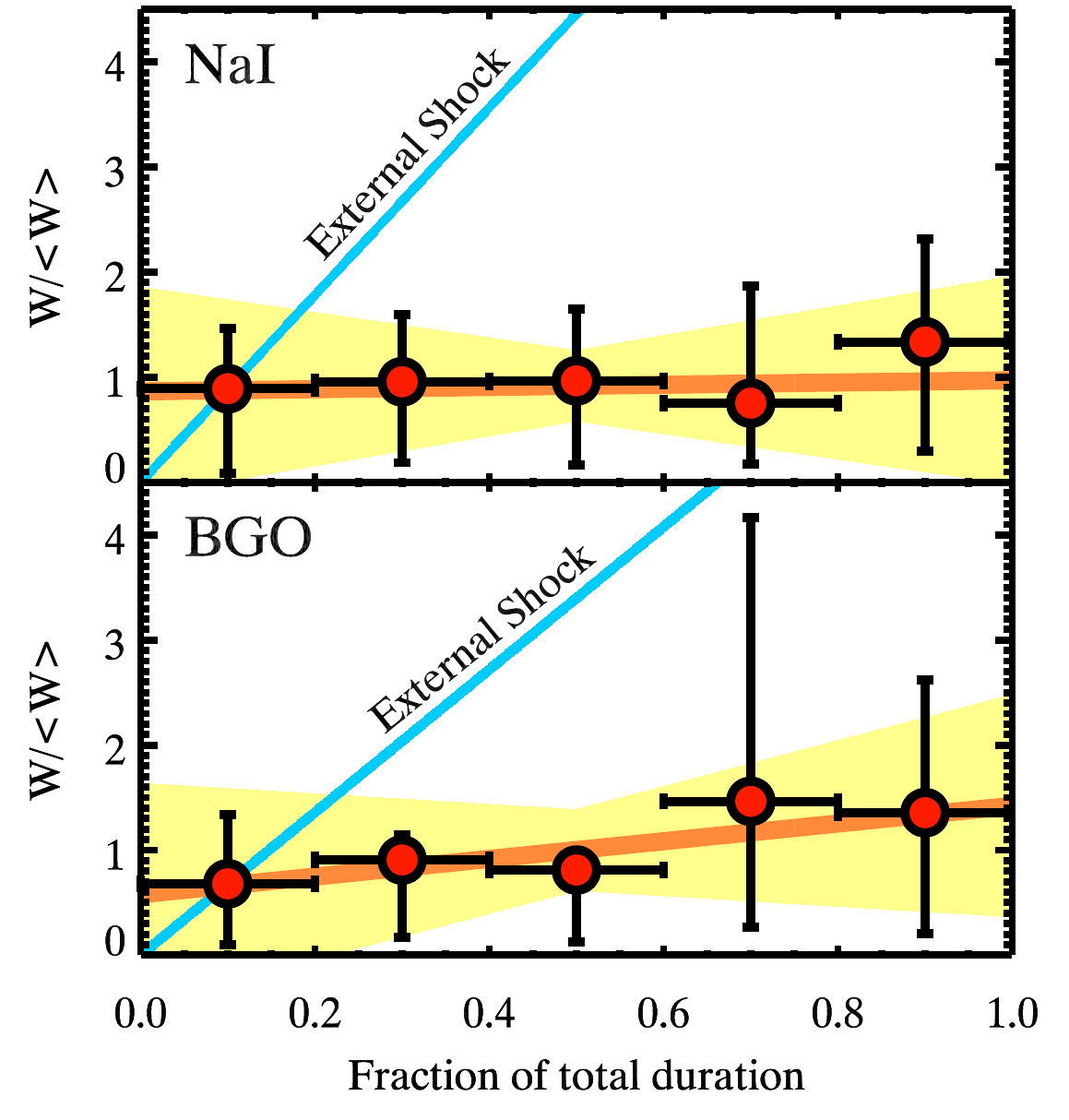}
\caption{\label{metodo2fig}{Average evolution of the normalised pulse width with time. The orange solid line represents the linear fit and the yellow shadow the associated uncertainty region. The blue line is the expected behavior assuming as emission mechanism external shocks happening at increasing distance. Top and bottom panels show the results obtained using the pulses of B12 obtained from the NaI and BGO \ba\ lightcurves respectively. }}
\end{figure}

In order to further support these results, we also compute the average pulse duration evolution along the GRB. 
To this aim, we study the average evolution of the normalized pulse width $W/\langle W \rangle$ of every single prompt pulse in B12 (purple squares and red diamonds in Fig.\ref{observer}). We divide each light curve into five sectors, which are fractions of the total GRB duration. For each sector we re--normalise every single pulse to the average duration of all the pulses belonging to the same GRB. So we averaged all the normalised pulses in the same sector. Fig.\ref{metodo2fig} shows the evolution with time of $W/\langle W \rangle$ for the NaI and BGO pulses. Errors represent the 68\% confidence interval of the normalised pulse width. 

We fitted, in the barycentre of the data points, a linear function $W/\langle W \rangle  = m t + q$ finding  $m=0.1 \pm 1.3 $, $q=-0.06 \pm 0.34 $ ($\chi ^2 = 0.24$) and $m=0.8 \pm 1.3 $ , $q=-0.04 \pm 0.39 $ ($\chi ^2 = 0.15$) for NaI and BGO pulses, respectively. The fits and their uncertainty are shown by the orange lines and yellow shaded regions in Fig.\ref{metodo2fig}. These results confirm that the pulse width remains constant with time. 
The average values of $W/\langle W \rangle$ of the five sectors and their errors are reported in Tab.\ref{tab1}. 

\begin{table}
\footnotesize
\centering
\begin{tabular}{ccccc}
\toprule
& NaI & & BGO & \\ 
Sector & mean $W/\langle W\rangle$ & \# pulses & mean $W/\langle W\rangle$ & \# pulses \\ 
\midrule
1  & $0.89_{-0.80}^{+0.57}$ & 78 & $0.68_{-0.58}^{+0.66}$ & 66  \\
2  &$0.96_{-0.77}^{+0.64}$ & 87  & $0.91_{-0.73}^{+0.24}$ & 54 \\
3  & $0.97_{-0.80}^{+0.68}$ & 74  & $0.81_{-0.69}^{+0.15}$ & 28 \\
4 & $0.75_{-0.58}^{+1.12}$ & 53  & $1.46_{-1.20}^{+2.70}$ & 28 \\
5 & $1.34_{-1.04}^{+0.98}$ & 82   & $1.36_{-1.15}^{+1.26}$ & 52 \\ 
\hline
\end{tabular}
\caption{\label{tab1}{Mean values of the normalized width $W/\langle W\rangle$ in each sector with associated superior and inferior estimated errors and relative number of pulses within the bin. Data refers to pulses coming from NaI and BGO light cuves.}}
\end{table}

\section{Internal vs External Shocks scenario}

The blue lines (normalised to the first data point) in Fig.\ref{metodo2fig} show how the pulse width should increase with time in the ES scenario (Eq. \ref{eq:4}). The absence of correlation between the pulse width and the peak time (shown by the correlation analysis presented in \S 2) and the almost constant average pulse width along the GRB light curves (as found in Fig.\ref{profilimedi}) favours IS as the leading mechanism for the origin of the dissipation associated with the $\gamma$--ray emission of GRBs. 

The linear increase of XRF duration with time (Fig.\ref{observer}) has been interpreted as a signature of their origin from external shocks generated by the interaction of the blast wave with shell--like over--dense regions located at increasing distance from the central engine \citep[see also][]{Wang2000ApJ...535..788W,Lazzati2002A&A...396L...5L,Heyl2003ApJ...586L..13H}. \cite{Chincarini2007ApJ...671.1903C} also discuss this mechanism considering the global fireball deceleration which could produce the superposition of flares with a monotonically decaying continuum emission (i.e. the standard afterglow).
 
\begin{figure}
\centering
\includegraphics[width= 0.5\textwidth]{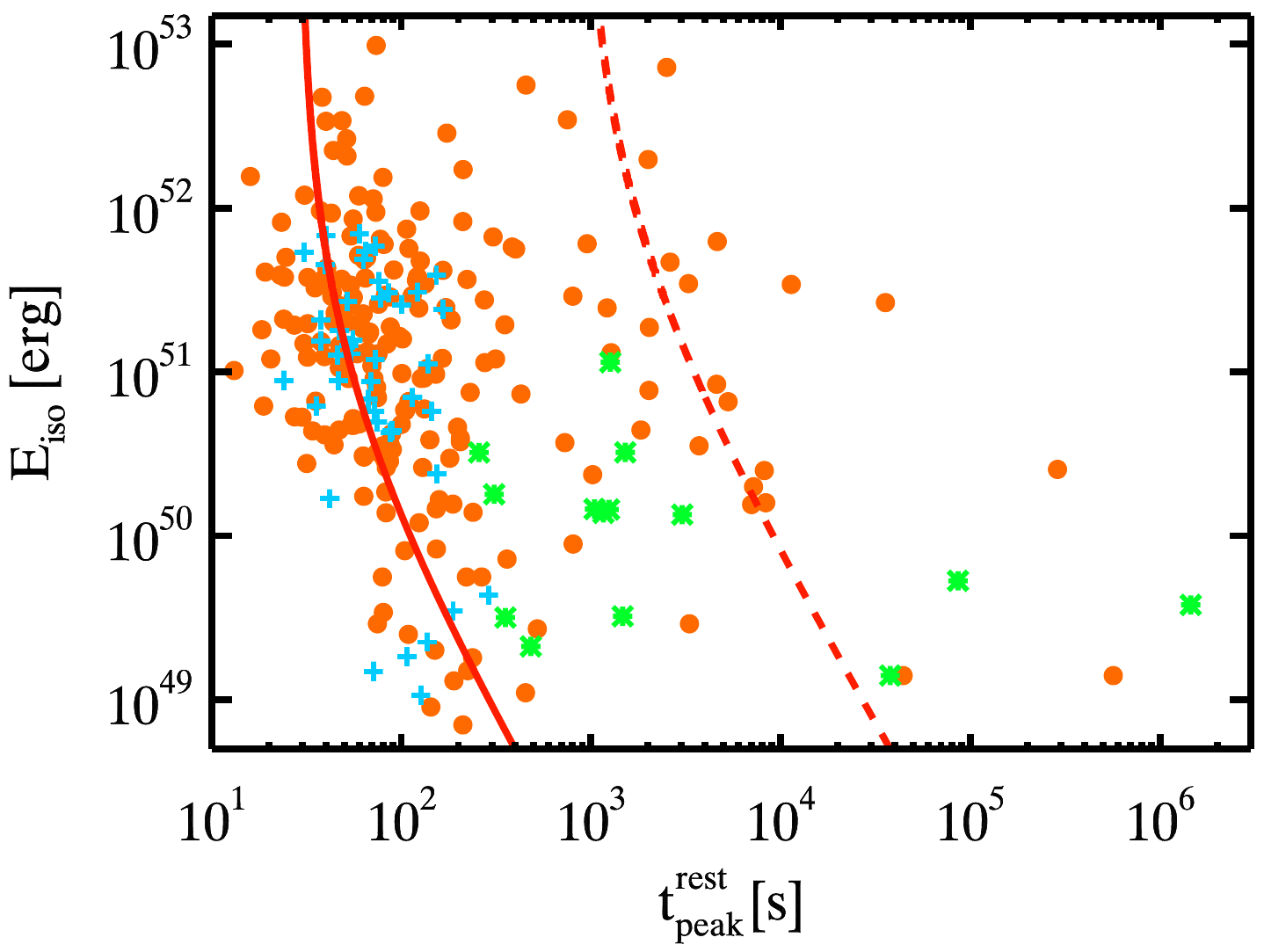}
\caption{\label{enevst}{Isotropic energy (computed in the XRT energy range) of XRF versus their rest frame peak time. Different symbols and color refer to different samples collected from the literature.  Orange points, green asterisks and cyan crosses are associated to X--ray Flares (only those ones having measured redshift) from \citeauthor{Yi2016ApJS..224...20Y}(\citeyear{Yi2016ApJS..224...20Y}), \citeauthor{Bernardini2011A&A...526A..27B}(\citeyear{Bernardini2011A&A...526A..27B}) and \citeauthor{Chincarini2010MNRAS.406.2113C}(\citeyear{Chincarini2010MNRAS.406.2113C}), respectively. The solid (dashed) red line shows the behaviour of the isotropic energy versus flare peak time predicted considering two shells emitted with a delay $\Delta T=30$ ($1000$) s with respect to the start of the prompt emission and with a separation $\Delta R=7\times10^{10}$ ($7\times10^{12}$) cm.}}
\end{figure}

The distribution of the PP and XRF in Fig.\ref{observer} suggests the presence of a continuous and monotonic trend. This motivates us to explore also the possibility that XRF can have an internal origin. The IS scenario predicts a constant pulse width and would apparently be disfavoured. However, for the IS case we assumed that the shock is produced at a constant distance from the central engine. If we relax this assumption, as we show in \S 2, it is possible to explain also XRF through IS produced by shells with slightly different bulk velocities. Additionally, another property of XRF which should be consistent with this interpretation is their energetic. Early flares show a possible decreasing isotropic equivalent energy as a function of their time of occurrence as pointed out by \cite{Margutti2011MNRAS.410.1064M}. We have computed the isotropic equivalent energy of  XRFs as $E_{\rm iso}=4 \pi d_{\rm L}^2 S / (1+z)$, where $S$ is the fluence in the XRT energy range. Fig.\ref{enevst} shows $E_{\rm iso}$ for XRFs with measured $z$ as a function of the rest frame peak time $t^{\rm rest}_{\rm peak}$ (different symbols refer to different samples - see \S 3). 

Fig.\ref{enevst} shows that there is a trend: flares occurring later are less energetic. We have investigated if this trend can be the result of a decreasing efficiency 
of internal shocks. Suppose that two shells are created at two different times, and have very similar velocities. The smaller the velocity difference, the longer it takes for them to collide, and the smaller the produced energy, because their relative kinetic energy is small. In this case we do expect a trend: the flare $E_{\rm iso}$ should decrease with $t_{\rm peak}^{\rm rest}$. In order to born this out, we derive how the energy, released during the flare, scales with the time of occurrence of the flare (Eq.\ref{eq:t_peak}). For the sake of simplicity we assume that the shell thickness is equal to their separation ($\delta R \simeq \Delta R$). Eq. \ref{eq:t_peak} becomes:
\begin{equation}\label{eq:t_peak2}
t_{\rm peak}^{\rm rest} \simeq \Delta T + \frac{\Delta R}{c}\frac{1}{\alpha_{\Gamma}-1}
\end{equation}

where $\Delta R$ represents the separation of the shells producing X--ray flares.

To evaluate the time of occurrence of flares $t_{\rm peak}^{\rm rest}$ we allow $\Delta T$, $\Delta R$ and $\alpha_{\Gamma}$ to vary in the other two terms.


An upper limit on the isotropic energy $E_{\rm iso}$ released in form of radiation during the flare can be derived applying the efficiency factor $\eta$ (Eq. \ref{eq:2}) to an initial isotropic equivalent kinetic energy $E_{\rm kin}$ (fixed at $10^{54}$ erg). Thus, knowing how the peak time and the efficiency vary with respect to $\alpha_{\Gamma}$ (fixing all other parameters), we can find how the emitted energy changes with respect to the time when the pulse occurs. Combining Eq.\ref{eq:2} and Eq.\ref{eq:t_peak2} we derived the analytic expression linking $E_{\rm iso}$ and $t_{\rm peak}^{\rm rest}$:
\begin{equation}
E_{\rm iso} = E_{\rm kin} \left\lbrace 1- \frac{2\sqrt{c(t_{\rm peak}^{\rm rest}-\Delta T)\left[ \Delta R + c(t_{\rm peak}^{\rm rest}-\Delta T \right]}}{\Delta R + 2c(t_{\rm peak}^{\rm rest}-\Delta T)} \right\rbrace
\end{equation}

Fig. \ref{enevst} shows the curves obtained assuming $\Delta R=7\times10^{10}$ ($7\times10^{12}$) cm and $\Delta T=30$ (1000) s as the solid (dashed) red line. Interestingly, independently from the parameter choice, for relatively small values of $\alpha_\Gamma$ (in the limit of shells with slightly different bulk Lorentz factors) the energy released in the flares scales as $t^{-2}$. These curves shows that for some particular combination of parameters, it is possible to produce flares and also to consistently produce an energy of the flares which decreases with the flare occurrence time. 

The same scenario allows us to derive the width of the flares as a function of their peak time. Assuming again $\delta R \simeq \Delta R$ from Eq. \ref{eq:delta} we obtain the rest frame pulse duration:
\begin{equation}
\Delta t^{\rm rest} = \frac{2\Delta R}{c}\frac{\alpha_{\Gamma}}{\alpha_{\Gamma}^2-1}
\end{equation}
Combining this equation with Eq.\ref{eq:t_peak2} we obtain the dependence of $\Delta t^{\rm rest}$ on the peak time:
\begin{equation}
\Delta t^{\rm rest} = 2(t_{\rm peak}^{\rm rest}-\Delta T) \frac{\Delta R + c(t_{\rm peak}^{\rm rest}-\Delta T)}{\Delta R + 2c(t_{\rm peak}^{\rm rest}-\Delta T)}
\end{equation}

\begin{figure}
\centering
\includegraphics[width= 0.5\textwidth]{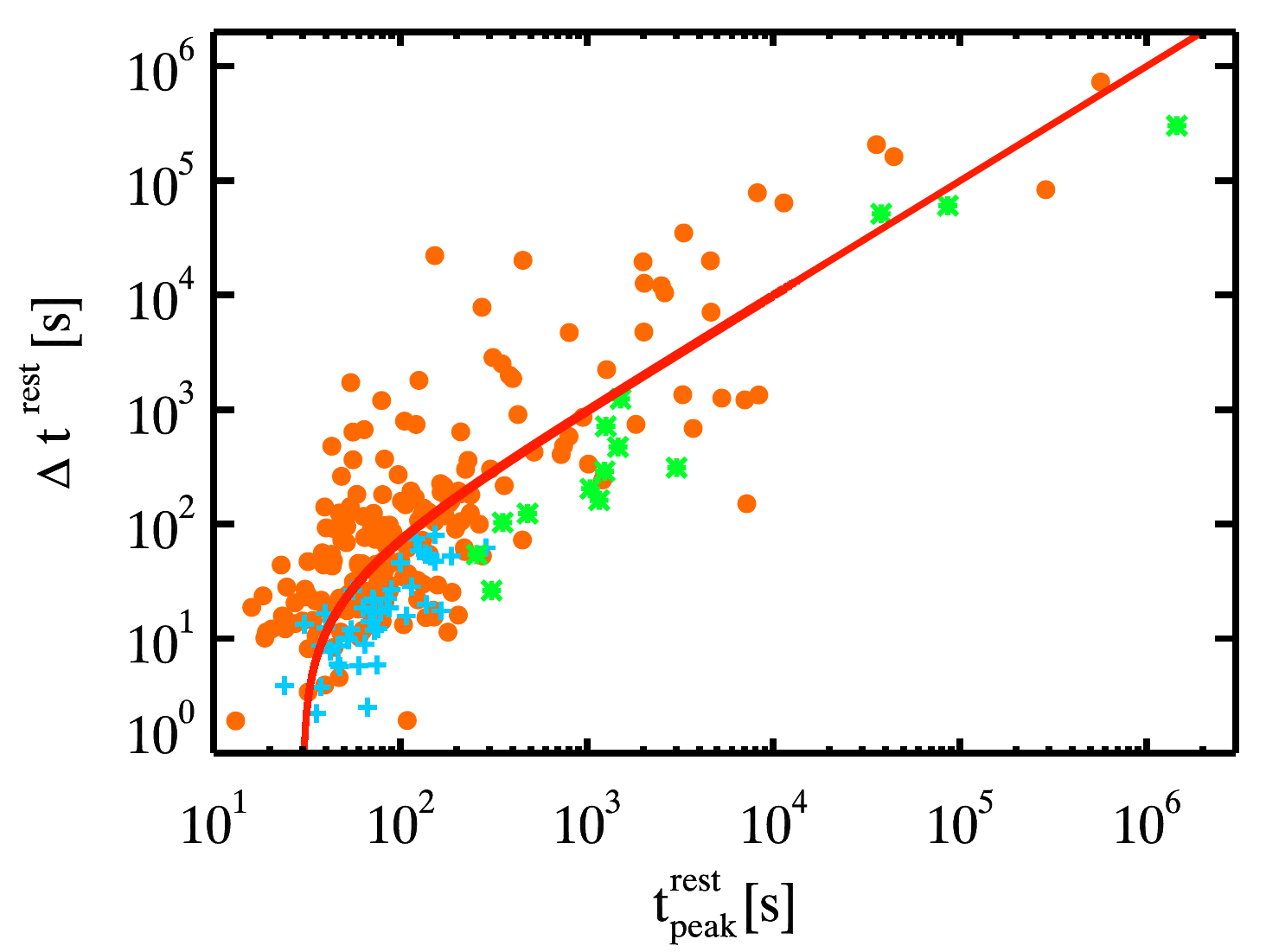}
\caption{\label{wvst}{Rest frame duration versus occurence time for XRFs.  Different symbols and color refer to different sources found in the literature for these parameters. Orange points, green asterisks and cyan crosses are associated to X--ray Flares (only those ones having measured redshift) from \citeauthor{Yi2016ApJS..224...20Y}(\citeyear{Yi2016ApJS..224...20Y}), \citeauthor{Bernardini2011A&A...526A..27B}(\citeyear{Bernardini2011A&A...526A..27B}) and \citeauthor{Chincarini2010MNRAS.406.2113C}(\citeyear{Chincarini2010MNRAS.406.2113C}), respectively. The red solid line shows the expected relation between the flare width and the time of the flare in the IS shock model (it is obtained adopting $\Delta T=30$ s and $\Delta R=7\times10^{10}$ cm.}}
\end{figure} 

This relation is shown by the red curve in Fig.\ref{wvst}: points show the data in the rest frame (for XRF with measured redshifts) and the curve is obtained assuming $\Delta R=7\times10^{10}$ cm and $\Delta T=30$ s. We are able to justify the increasing duration of the flares with time. 
The leading dependence of the flare properties (temporal and energetic) is from $\alpha_\Gamma$. The smaller is $\alpha_\Gamma$ (i.e. shells with only slightly different bulk Lorentz factors) the larger is the flare duration and its occurrence time and the smaller is the released energy.  
With this simple model we do not aim to derive the parameters of the shells which is out of the scopes of the present work. Our simple approach considers only two shells encounters. It can explain the overall temporal and energetic tendency of the majority of flares as a succession of IS with decreasing values of $\alpha_\Gamma$. Allowing for some dispersion of free parameters it might be possible to account for the dispersion of the XRF in  Fig.\ref{enevst} and Fig.\ref{wvst}. Very late flares could still be explained choosing a particular set of parameters (maybe extreme values) so it seems reasonable also to consider the possibility that their origin may be different.

\section{Discussion \& Conclusion}
\label{Conclusions}

In this work we considered the possible dependence of the pulse width with time along GRBs combining prompt emission pulses with X--ray flares. It is known that prompt emission pulses do not show an increase of the pulse width with time \citep{Ramirez2000ApJ...539..712R}. XRF, instead, show a nearly linear increase of their width with time \citep{Chincarini2010MNRAS.406.2113C,Yi2016ApJS..224...20Y,Kocevski2007ApJ...667.1024K}. Fig.\ref{observer} shows prompt emission pulses (from the sample of B12) and XRFs (from the samples of C10, B11, Y16). 

Internal shocks predict that the dissipation of energy between shells coasting with slightly different bulk Lorentz factor should produce random pulses with duration uncorrelated with their time of occurrence. On the other hand, the flare width has been observed to increase with time and considered so far a signature in favour of an external origin. In this scenario, the production of flares occurs in the interaction of a decelerating blast wave with ISM over--densities located at increasing distance from the central engine. In these two scenarios, what determines whether the pulse duration increases with time or not is the dissipation at increasing radii (as in external shocks) or at a constant distance (as in internal shocks). 

We verified with three different methods (Fig.\ref{observer} - bottom--right inset, Fig.\ref{profilimedi} and Fig.\ref{metodo2fig}) and with two independent samples of prompt emission pulses (pulses obtained by the deconvolution of bright \fe\ bursts - from B12 - and a sample of 100 GRBs with redshift whose light curves were analyzed in this work) that prompt emission pulses show no correlation between their duration and their time of occurrence. Our results fully confirm those obtained with the BATSE data by \cite{Ramirez2000ApJ...539..712R}. Despite the emission during the prompt phase is highly variable, it can be described as the emission due to internal shocks occurring almost at the same distance from the central engine. 

XRFs exhibit an (almost linear) increase of their duration with time. Moreover, their peak luminosity $L_{\rm p}$ anti-correlates with the peak time $t_{\rm peak}$: for early flares \citep[$t_{\rm peak} < 1000$ s,][]{Chincarini2010MNRAS.406.2113C,Margutti2011MNRAS.410.1064M}  $L_{\rm p} \propto t^{-2.7}$ and becomes shallower $L_{\rm p} \propto t^{-1.7}$ for late flares \citep{Bernardini2011A&A...526A..27B}. As a consequence, \citep[as noted by][]{Margutti2011MNRAS.410.1064M} the energy released during the flares should scale as $t^{-1.7}$ ($t^{-0.7}$ for late time flares).

The origin of XRF has been debated in the literature. Temporal and spectral properties of XRFs lead different authors to ascribe them to internal--like dissipation due to the late time activity of the inner engine \citep{Falcone2006ApJ...641.1010F,Falcone2007ApJ...671.1921F,Lazzati2007MNRAS.375L..46L,Maxham2009ApJ...707.1623M}.
Alternatively, XRFs could be produced by external shocks with over--dense regions of the ISM \citep[e.g.][]{Wang2000ApJ...535..788W,Lazzati2002A&A...396L...5L,Heyl2003ApJ...586L..13H,Nakar2007MNRAS.380.1744N} or by the long--lived reverse shock interacting with the tail of the ejecta \citep{Hascoet2015arXiv150308333H}.

Based on the distribution of prompt emission pulses and XRF in the plane of Fig.\ref{observer} we considered the possibility that also XRF are produced by internal shocks between shells emitted with a certain initial separation and a certain (even small) temporal delay with respect the prompt shell. If these shells are characterized by small values of $\alpha_{\Gamma}$, the time of their encounter is delayed (and therefore the shock development). Later flares last longer and are less efficient in emitting radiation. These results, shown by the red model curves in Fig.\ref{enevst} and Fig.\ref{wvst}, are consistent with the distribution of data in these planes. In this scenario the leading parameter is the relatively low ratio between the shells' Lorentz factors, parametrized by $\alpha_\Gamma$. The asymptotic behaviour, for small $\alpha_\Gamma$, is approximately $t^{-2}$ in agreement with the $E \propto L_{\rm p} \Delta t \propto t^{-1.7}$ also marginally shown by the early flares in Fig.\ref{enevst}. This behavior also seems to explain the shape of the left boundary of the distribution in the region populated by XRF with low energies and peak time.
 


We showed that it is possible to explain the energetic and temporal properties of X--ray flares as the result (under appropriate assumptions) of "classical" internal shocks between fireball ejected during the prompt emission phase. We do not require that the inner engine is active until late times:  late flares, characterized by smaller energies, can be produced by relativistic shocks between fireballs with Lorentz factor ratio $\alpha_{\Gamma}$ very close to one. 

Our results pose the question on the mechanism which is responsible for having pairs of shells
with largely different bulk Lorentz factors (large $\Gamma$--contrasts) early on, producing the prompt emission, 
and pairs of shells with smaller $\Gamma$--contrast later, producing the X--ray flares.
While the answer to this question is out of the scope of the present paper, one possible idea has been suggested \citep{Ghisellini2007ApJ...658L..75G,Ghisellini2009MNRAS.393..253G}, suggesting that there can be two phases of accretion onto the black hole.
The first is the accretion of the very dense material left over by the collapse of the core of the star.
This dense material can sustain very large magnetic fields that can extract the spin energy of the black hole.
The power so extracted is very large allowing the formation of shells of very large $\Gamma$--factors.
If the accretion is modulated, or quasi--intermittent, then it is possible to form shells with very different
energetics and bulk Lorentz factors. 
This phase is followed by the accretion of fallback material, less dense. 
This corresponds to the extraction of less spin energy from the black hole, and presumably 
both the maximum and the average values of the bulk Lorentz factors are smaller, as well as
the $\Gamma$--contrast between consecutive shells.

	




\section*{Acknowledgments}
We thank the referee for the valuable comments and suggestions that helped us improving the manuscript. We acknowledge 2014 PRIN-INAF grant for financial support. MR thanks the Brera Observatory for the kind hospitality during his graduate work.

\bibliographystyle{aa}
\bibliography{journals,bibliografia}

\label{lastpage}
\end{document}